# Two-dimensional superconductivity at (110) LaAlO$_3$/SrTiO$_3$ interfaces

Yin-Long Han[1,§], Sheng-Chun Shen[1,§], Jie You[2,§], Hai-Ou Li[2], Zhong-Zhong Luo[1], Cheng-Jian Li[1], Guo-Liang Qu[1], Chang-Min Xiong[1], Rui-Fen Dou[1], Lin He[1], Donald G Naugle[3], Guo-Ping Guo[2,*] and Jia-Cai Nie[1,†]

[1] Department of Physics, Beijing Normal University, Beijing, 100875, People's Republic of China
[2] Key Laboratory of Quantum Information, Department of Optics and Optical Engineering, University of Science and Technology of China, Chinese Academy of Science, Hefei 230026, People's Republic of China
[3] Department of Physics, Texas A&M University, College Station, TX 77843, USA

Novel low dimensional quantum phenomena are expected at (110) LaAlO$_3$/SrTiO$_3$ (LAO/STO) interfaces after the quasi two dimensional electron gas similar to that of (001) LAO/STO interfaces was found [G. Herranz et al., Sci. Rep. **2**, 758 (2012) and A. Annadi et al., Nat. Commun. **4**, 1838 (2013)]. Here, two dimensional superconductivity of (110) LAO/STO samples with a superconducting transition temperature of ≅ 184 mK is demonstrated based on systematical transport measurements. The two dimensional characteristic of the superconductivity is confirmed by analyzing the Berezinskii-Kosterlitz-Thouless transition. The estimated superconductive thickness is about 18 nm. These features of superconductivity of (110) LAO/STO interfaces are comparable to those of (001) LAO/STO interfaces. This discovery may inspire a new round of upsurge on study of LAO/STO interfaces.

## I. INTRODUCTION

Due to the strongly correlated characteristics of electrons, transition metal oxides (TMO) have attracted extensive attention of the researchers. Especially the heterojunctions of TMO, benefiting from the enhancement of the interplay between the charge, spin, orbital, and lattice degrees of freedom, exhibit a variety of novel phenomena not found in the respective bulk constituents or in conventional semiconductor interfaces. One of the most fascinating systems is (001) LaAlO$_3$/SrTiO$_3$ (LAO/STO) interfaces, at which high mobility quasi two dimensional electron gas (2DEG),[1–4] two dimensional (2D) interface superconductivity,[5–7] ferromagnetism,[8–13] and the coexistence of superconductivity and ferromagnetism[14–16] have been observed. Therefore, the research of experiment and theory of (001) LAO/STO interfaces has become one of the hotspots in the research fields of TMO heterojunctions. Recently, at (110) orientated LAO/STO interfaces, unexpected high mobility metallic conductivity was also found.[17,18] The quasi 2DEG of the (110) LAO/STO interfaces has similar values of the sheet carrier density, room temperature resistivity and electronic mobility to those found at the (001) LAO/STO interfaces. In addition, a unified mechanism of the origin of the 2DEG for both (001) and (110) LAO/STO interfaces has been proposed.[19] So the fascinating novel phenomena resided at the (001) LAO/STO interfaces is also expected at the new (110) LAO/STO interfaces. What is more, because there are two different crystal orientations, i.e., (010) and (1-10), in the plan of (110) LAO/STO interfaces, the added degree freedom of crystal anisotropy would play a role and may produce newfangled physics.[20–22] Therefore, the discovery of the quasi 2DEG at (110) LAO/STO interfaces injects new vigor and presents new opportunities into the research of heterojunctions of TMO. While the deep study on (110) LAO/STO interfaces is lack, particularly, the ground state, whether the electrons condensed into a ferromagnetic state or a superconducting state when the temperature approaches absolute zero, has not been reported by now.

In this study, we demonstrate that the electrons at the investigated (110) LAO/STO interfaces condensed into a superconducting ground state. The superconducting transition temperature ($T_c$) is about 184 mK. The characteristics of the superconductive transition are consistent with a Berezinskii-Kosterlitz-Thouless (BKT) transition[23–25] of a 2D electron system which proves the superconductivity is 2D of the interface but not of the bulk STO substrate. The BKT transition temperature ($T_{BKT}$) is about 177 mK. The estimated thickness of 2D superconducting layer is about 18 nm.

## II. EXPERIMENT

In this study, the films with 5 unit cells (uc) of LAO was grown by pulsed laser deposition (PLD) at (110) STO substrates. The as-received (110) STO substrates were treated at 1050 °C for 2h under oxygen atmosphere,[26] atomically flat (110) STO surfaces were obtained. The substrate was heated from room temperature to 750 °C in 0.1 mbar of O$_2$, and then the LAO layer was grown in 10$^{-5}$ mbar of O$_2$. The laser pulses were supplied by a KrF excimer sources (λ=248nm) with an energy density of 1.5 J/cm$^2$ and a frequency of 1 Hz. After deposition, every sample was cooled down in oxygen rich atmosphere to avoid the formation of oxygen vacancies:[17,27,28] P$_{O2}$ = 0.3 mbar from 750 °C to 600 °C and P$_{O2}$ = 200 mbar from 600 °C to room temperature, with a dwell time of 1 h at 400 °C.

The measurements of the sheet resistance ($R_s$) and voltage versus current (V-I) curves were performed by using four contacts arranged on the samples as shown in the upper insert of Fig. 1(a). The contacts were formed by ultrasonic welding using 25-micrometer-diameter Al-wires directly connected to the sample interfaces. The measurements were mainly performed in a dilution refrigerator.

## III. RESULTS AND DISCUSSION

The typical transport properties of the 5-uc (110) LAO/STO samples are shown in Fig. 1. The temperature ($T$) dependence of resistance ($R$) exhibits clear metallic behavior, as shown in the lower insert of Fig. 1(a). There is an obvious superconducting transition with a $T_c \cong 184$ mK [Fig. 1(a)]. The $T_c$ was defined as the temperature where the resistance is the 50% of its normal state value ($R_n$, measured at T = 400 mK). The width of the transition $\Delta T_c$ (defined between 20% and 80% of $R_n$) is $\cong 36$ mK. The $T_c$ and $\Delta T_c$ of the (110) LAO/STO interfaces here are comparable to those of superconductive (001) LAO/STO interfaces[5,6] as well as (001) LaTiO$_3$/STO interfaces.[29,30] The $I$-$V$ curve at $T = 50$ mK in zero magnetic field indicates a supercurrent lasting up to ~ 14 μA [Fig. 1(b)]. Furthermore, as seen in Fig. 1(c), at $T = 50$ mK, when a magnetic field perpendicular to the sample surface ($H_\perp$) is applied, $R_s$ recovers its normal-state value above 0.4 T, while, when the magnetic field is parallel to the surface ($H_\parallel$), $R_s$ recovers its normal-state value above 2 T. The $R_s$-$H$ curves exhibit obvious anisotropy on the orientation of magnetic field. The occurrence of the zero-resistance state and the characteristic of both $I$-$V$ and $R_s$-$H$ dependencies clearly demonstrate superconductivity at the (110) LAO/STO interfaces. The electric current is perpendicular to the magnetic field during the measurements for $H_\parallel$ case.

The obvious anisotropy of $R_s$-$H$ curves at $H_\perp$ and $H_\parallel$ [Fig. 1(c)] indicates that the superconductivity in our (110) LAO/STO interfaces should be 2D characteristic. For a 2D system, it is well established that the superconducting transition should obey a BKT transition, characterized by a transition temperature $T_{BKT}$.[31–36] Below $T_{BKT}$, vortices and antivortices pairs are formed, and the zero-resistance state appears. As temperature rising, a thermodynamic instability should occur in which vortex-antivortex pairs, bound at low temperatures, spontaneously unbind into free vortices at the characteristic transition temperature $T_{BKT}$, and the proliferation of free vortices destroys superconductivity. According to the BKT transition theory, near the $T_{BKT}$, (I) a strong non-Ohmic behavior in the $V$-$I$ characteristics displays, in a simple power law $V \propto I^{a(T)}$ with $a = 3$ at $T_{BKT}$;[32,35,37] and (II) the temperature dependence of resistance behaves as $R(T) = R_0 e^{-b(T-T_{BKT})^{-1/2}}$,[32,36] which generates $[d\ln R(T)/dT]^{-2/3} = (2/b)^{2/3}(T - T_{BKT})$, i.e., $[d\ln R(T)/dT]^{-2/3}$ is linear with $T$. The above two features have been commonly used to experimentally demonstrate the existence of a BKT transition and to determine the value of the $T_{BKT}$.[5,36–40]

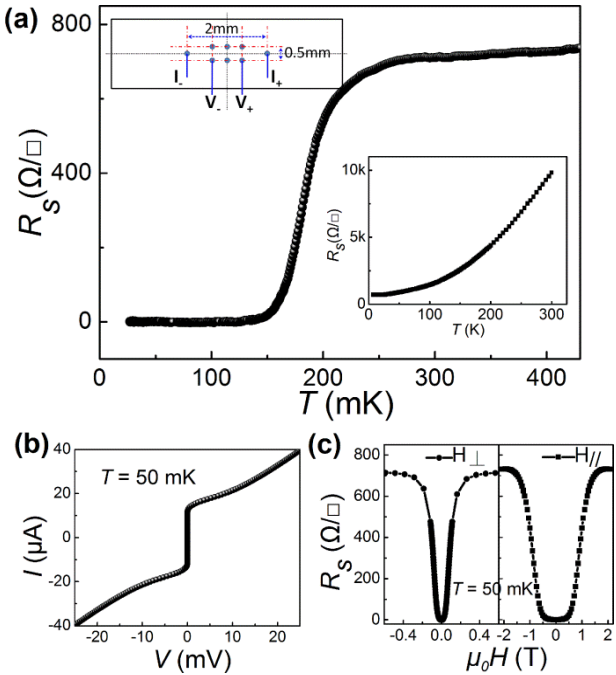

**FIG. 1** (Color online) (a) Dependence of the sheet resistance on $T$ of the (110) LaAlO$_3$/SrTiO$_3$ sample with 5-uc LaAlO$_3$ overlayers (measured with a 50-nA current). The upper inset shows the arrangement of the electrodes for four-terminal transport measurements. The lower inset shows sheet resistance versus temperature measured between 8 K and 300 K. (b) The $I$-$V$ curve at 0.05 K. (c) The the magnetic-field ($\mu_0 H$) dependence of $R$s at 0.05 K. $H_\perp$ and $H_\parallel$ mean the magnetic field was applied perpendicular and parallel to the sample surface, respectively.

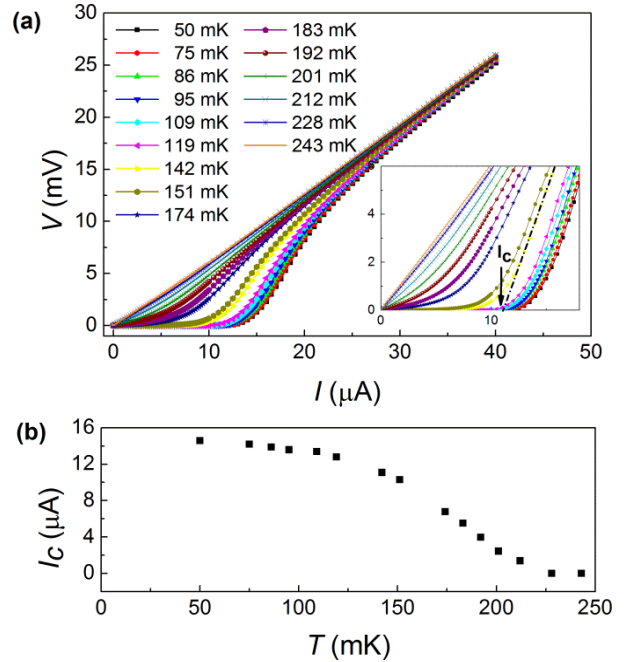

**FIG.2** (Color online) (a) Temperature-dependent $V$-$I$ characteristics of the 5-uc (110) LaAlO$_3$/SrTiO$_3$ interfaces measured at different temperature. The insert shows that the critical current ($I_c$) was gained by extrapolating the linear part near the upward corner of the $V$-$I$ curves to zero axis. (b) Temperature dependence of the $I_c$ obtained from (a).

Therefore, in order to examine the 2D superconductive characteristics of our 5-uc (110) LAO/STO heterojunction samples, we measured the *V-I* characteristics from 50 mK, where the samples are completely superconductive, to higher temperature where the samples recover to normal state. The representative results of a sample same to that of Fig. 1 are shown in Fig. 2(a). At 50 mK, for small current, there is a clear superconductive terrace at the *V-I* curve, and as the current increases, the *V-I* curve recovers to a linear Ohmic characteristic in the end. With the increase of the temperature, the superconductive terrace becomes shorter till vanishes, and lastly, the *V-I* turn into almost complete linear Ohmic characteristic at 243 mK. The critical current ($I_c$) of each temperature was defined by extrapolating the linear part near the upward corner of the *V-I* curves to zero axis,[41] as shown in the insert of Fig. 2(a). The temperature dependence of $I_c$ is shown in Fig. 2(b), the $I_c$ decreases from ~ 14 μA at 50 mK to zero at 228 mK and 243 mK, exhibiting a steplike structure similar to that of (001) LAO/STO interfaces.[5] The above *V-I* characteristics at different temperature provide strong evidence for the superconductivity of the (110) LAO/STO interfaces. More importantly, we replot the *V-I* data of Fig. 1(a) in a logarithmic scale to check whether the superconductive transition of the (110) LAO/STO interfaces here is consistent with a BKT transition. As indicated by short black lines in Fig. 3(a), near the $T_c$ ($\cong$ 184 mK), the *V-I* curves indeed show a clear V $\propto$ $I^{a(T)}$ power-law dependence. So the (110) LAO/STO samples does undergo a BKT transition according to the feature (I) of the BKT transition. The long black line corresponds to $V \sim I^3$ dependencies and shows that 174 mK < $T_{BKT}$ < 183 mK. The power exponents $a(T)$ are gained by fitting the data and plotted as a function of temperature [Fig. 3(b)], revealing that the $T_{BKT} \cong 177$ mK.

Furthermore, the *R-T* characteristics should also be consistent with a BKT transition. As expected by the feature (II) of the BKT transition, near the $T_{BKT}$, $[dlnR(T)/dT]^{-2/3}$ should be linear with *T*. As shown by a blue line in Fig. 3(c), the $[dlnR(T)/dT]^{-2/3}$ does scale linearly with *T* near 200 mK. The $T_{BKT}$ can be extracted from the intersection points of the linear part with the zero axis.[5,36] The $T_{BKT}$ extracted by this method is $\cong$ 182 mK, in line with the the result of *V-I* analysis.

To sum up, the characteristics of the superconductive transition of the (110) LAO/STO samples are consistent with the BKT transition, providing the powerful proofs of the 2D superconducting of the interfaces not of the bulk STO substrates. Similar results had been reported at (001) LAO/STO interfaces.[5,6]

For purpose of further investigating the 2D character of the superconductive state, we investigated the *V-I* characteristics of the samples at 50 mK for different values of magnetic fields applied both perpendicular and parallel to the interface. As illustrated in Fig. 4(a) and (d) for $H_\perp$ and $H_\parallel$, respectively, the *V-I* curves progressively recover to a linear Ohmic characteristic indicating the superconductive states are progressively suppressed as the magnetic field increases. However, it is very clear that the response of the samples to the magnetic field exhibits very strong dependence on the magnetic field orientation. It is 145 mT at which the superconductive state is almost suppressed for $H_\perp$ case [Fig. 4(a)], while the corresponding value is 1500 mT for $H_\parallel$ case [Fig. 4(d)], one order of magnitude larger than that of $H_\perp$. This strong anisotropy again proves the low-dimensional superconductivity of the samples, since the 2D superconductivity cannot be suppressed by vortex entry for in-plane fields,[42] and is in consistent with the BKT transition analysis (Fig. 3) and the $R_s$-*H* characteristics [Fig. 1(c)].

For more in-depth quantitative analysis, we get the differential resistance *dV/dI* directly deduced from the *V-I* data. The values of *dV/dI* represent the resistance of the sample at different measuring current. As shown in Fig. 4(b) and (e), for zero and lower magnetic field, the values of *dV/dI* are zero at lower measuring current, corresponding to the superconductive state, as the measuring current rises, the *dV/dI* increases to a peak value, and then decreases to a saturation $(dV/dI)_{I=40}$ at *I* = 40 μA at last, corresponding to the normal state; for higher magnetic, the *dV/dI* increases from not zero but a finite value $(dV/dI)_{I=0}$ at *I* = 0 μA, and the higher of the magnetic field, the bigger of the $(dV/dI)_{I=0}$. To elucidate this more clearly, as shown in Fig. 4(c) and (f), we plot the magnetic field $\mu_0 H$ dependence of the ratio $(dV/dI)_{I=0}/(dV/dI)_{I=40}$. Here, we use an innovative method, basing on the *V-I* characteristics of the samples at different values of magnetic fields, to determine the upper critical field ($H_{c2}$). The field at which the resistance at *I* = 0 μA recovers to half that of the normal value, *i.e.*, (d*V*/d*I*)$_{I=0}$/(d*V*/d*I*)$_{I=40}$ = 0.5, was define as the $H_{c2,50mK}$. As shown in Fig. 4 (c) and (e), we get $H_{c2\perp,50mK} \cong$ 104 mT and $H_{c2\parallel,50mK} \cong$ 1024 mT for $H_\perp$ and $H_\parallel$, respectively. According to the Ginzburg–Landau theory,

$$H_{c2,T} = H_{c2,0} \frac{1-[T/T_c(0)]^2}{1+[T/T_c(0)]^2}, \quad (1)$$

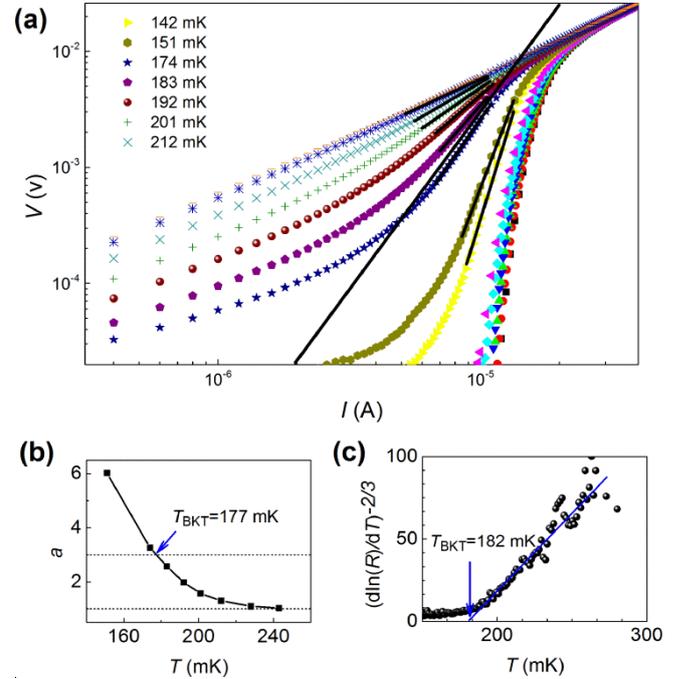

**FIG.3** (Color online) (a) Temperature-dependent *V-I* curves of Fig. 2(a) on a logarithmic scale. The color code is the same as that in Fig. 2(a). The short black lines are fits of the data in the transition. The long black line corresponds to $V \sim I^3$ dependencies and show that 174 mK < $T_{BKT}$ < 183 mK. (b) Temperature dependence of the power-law exponent a, deduced from the fits shown in (a). $T_{BKT}$ = 177 mK was deduced. (c) *R*(T) dependence of the 5-uc (110) LaAlO$_3$/SrTiO$_3$ sample, plotted on a [dln(*R*)/d*T*]$^{-2/3}$ scale. The blue solid line indicates the linear behavior expected for a BKT transition with $T_{BKT}$ =182 mK.

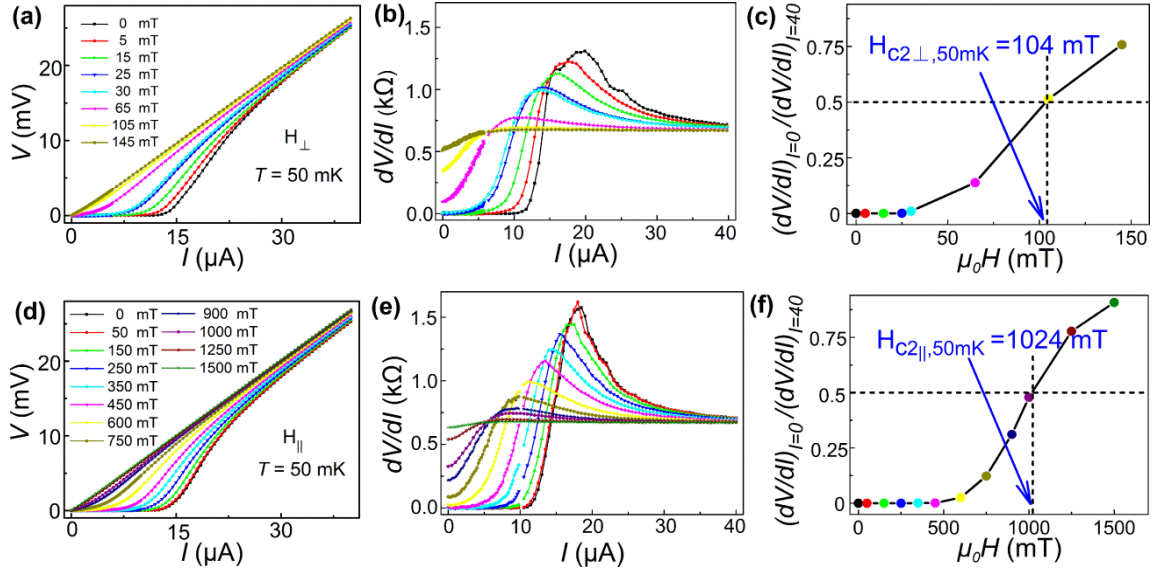

**FIG.4** (Color online) (a) Magnetic field-dependent *V-I* characteristics of the 5-uc (110) LaAlO$_3$/SrTiO$_3$ heterostructures at 0.05K for different magnetic field applied perpendicular (H$_\perp$) to the interface. (b) Numerical d*V*/d*I-I* curves deduced from the *V-I* curves in (a). (c) shows the (d*V*/d*I*)$_{I=0}$/(d*V*/d*I*)$_{I=40}$ at different perpendicular field. The filed (104mT) at which (d*V*/d*I*)$_{I=0}$/(d*V*/d*I*)$_{I=40}$ = 0.5 was defined as the $H_{c2\perp,50mK}$. (d), (e) and (f) The corresponding results for different magnetic field applied parallel (H$_\parallel$) to the interface. The H$_{c2\parallel,50mK}$ is 1024mT.

Where $T_c(0)$ is $T_c$ with no magnetic field, $T_c(0) \cong$ 184 mK here, $H_{c2,T}$ is the the upper critical field at a specific temperature $T$, $H_{c2,0}$ is the the upper critical field at 0 K. So we can deduce the $H_{c2\perp,0K} \cong$ 120 mT and $H_{c2\parallel,0K} \cong$ 1187 mT by using Eq. (1). More importantly, based on the values of $H_{c2\perp,0K}$ and $H_{c2\parallel,0K}$, a quantitative estimation of both the superconductive layer thickness $d$ and the in-plane superconductive coherence length $\varepsilon$ is feasible. Using a mean field approach leads to an in-plane coherence length[7] $\varepsilon = (\Phi_0/2\pi\mu_0 H_{c2\perp,0K})^{1/2} =$ 52 nm at $T$ = 0 K, $\Phi_0$ is the flux quantum. Then, according to Ginzburg–Landau theory[7] $d = (\sqrt{3}\Phi_0)/(\pi\varepsilon\mu_0 H_{c2\parallel,0K})$ = 18 nm. The estimated superconductive thickness $d$ is very smaller than the in-plane coherence length $\varepsilon$, well proves the 2D superconductivity of the (110) LAO/STO samples. The method to determine $d$ and $\varepsilon$ here had been widely used for both (001) LAO/STO interfaces and other STO based interfacial superconductors.[7,25,39,40] Although Ref. 43 believed that the value of $d$ by this method is the thickness upper limit, this estimated $d$ here still could represent the characteristic thickness of the 2D superconductivity.

The violations of the Pauli paramagnetic limit at both (001) LAO/STO interfaces and other STO based 2D superconductors are very common,[7,43,44] especially when the superconductive thickness $d$ < 20 nm. Similarly, the paramagnetic limit seems to be exceeded by a factor of three at (110) LAO/STO samples here. The Pauli paramagnetic limiting field is $\mu_0 H_c^P = 1.76 k_B T_c/(\sqrt{2}\mu_B)$, where $\mu_B$ is the Bohr magneton (with a $g$ factor of 2), $k_B$ is Boltzmann's constant, and $T_c \cong$ 0.184 K. So the $\mu_0 H_c^P =$ 340 mT, while the $H_{c2\parallel,0K} \cong$ 1187 mT.

Given the difference between the Ti 3$d$ subbands of (110) LAO/STO interfaces and those of (001) LAO/STO ones,[20–22] *e.g.*, the Ti 3$d_{xz}/d_{yz}$ orbitals lie at the lowest energy rather than the Ti 3$d_{xy}$ for (110) LAO/STO,[21,22,45] while which is the opposite for (001) LAO/STO,[20,46–48] it is beyond expectation that these characteristics of 2D superconductivity at the (110) LAO/STO interfaces, *i.e.*, the values of $T_c$, $T_{BKT}$, the in-plane coherence length and the superconductive thickness, and the violations of the Pauli paramagnetic limit, are all comparable to those of (001) LAO/STO interfaces. In spite of the origin of the superconductivity of (001) LAO/STO interfaces, also that of (110) LAO/STO interfaces, is not fully understood, the unexpected similarity suggests that the ground state and the mechanism of the superconductivity at both (001) and (110) LAO/STO interfaces should be same. Other layered superconductors, *e.g.*, YBa$_2$Cu$_3$O$_{7-x}$, MgB$_2$, and iron-based superconductors, exhibit anisotropic electronic properties due to the anisotropic crystal structure, while the situation in STO-based superconductors is different. STO is virtually cubic at low temperatures, so it is not surprising that the similarity of the superconductivity at (001) and (110) LAO/STO interfaces. More details about the mechanism of the superconductivity at both (001) and (110) LAO/STO interfaces need further deep studies.

Although having been investigated maturely about (001) LAO/STO interfaces, the origin of the condensed 2D electron system is still under debate by now. Relatively, the study of (110) LAO/STO interfaces is in its infancy, much more attention is needed to verify the other expected phenomena similar to those of the (001) LAO/STO interfaces, such as ferromagnetic order, electric field tunable superconductor-insulator quantum phase transition and so on, and as well as to explore the stimulated new novel physics different to those of (001) LAO/STO interfaces due to the unique Ti 3$d$ subbands[20–22] when added one more degree freedom of crystal anisotroy at (110) LAO/STO interfaces. The similarities of the superconductivity of (001) and (110) LAO/STO interfaces also imply that the condensed quasi 2DEG at both (001) and (110) LAO/STO interfaces may have a unified origin. Therefore, further in-depth investigation of (110) LAO/STO interface is very necessary and important which could provide great help in understanding the mechanisms of the novel phenomena at both (001) and (110) LAO/STO interfaces.

## IV. CONCLUSIONS

In summary, by systematical transport measurements, we demonstrate that the 5-uc (110) LAO/STO interfaces exhibit a 2D superconductivity with a $T_c \cong 184$ mK. The characteristics of the superconductive transition are consistent with a BKT transition. The BKT transition temperature is about 177 mK. The estimated thickness of the 2D superconducting layer is about 18 nm. The Pauli paramagnetic limit was exceeded by a factor of three. These features of superconductivity of the (110) LAO/STO interfaces are comparable to those of (001) LAO/STO interfaces suggesting that the ground state and the mechanism of the superconductivity at both (001) and (110) LAO/STO interfaces should be same.

## ACKNOWLEDGEMENTS

This work was supported by the Ministry of Science and Technology of China (Grant Nos. 2011CBA00200, 2013CB921701, 2013CBA01603, 2014CB920903), and the National Natural Science Foundation of China (Grant Nos. 10974019, 51172029, 91121012, 11004010, 61125403, 11374035, 11474022, 11222438, 61306150, 91121014 and 11311120461), the program for New Century Excellent Talents in University of the Ministry of Education of China (Grant No. NCET-13-0054), and Beijing Higher Education Young Elite Teacher Project (Grant No. YETP0238). The Texas A&M contribution was supported by TAMU-NSFC (Proj. 2014-030) and the Robert A. Welch Foundation, Houston, TX (Grant A-0514).

§These authors contributed equally to this paper.
*gpguo@ustc.edu.cn
†jcnie@bnu.edu.cn

[1] A. Ohtomo and H.Y. Hwang, Nature **427**, 423 (2004).
[2] M. Huijben, G. Rijnders, D.H.A. Blank, S. Bals, S.V. Aert, J. Verbeeck, G.V. Tendeloo, A. Brinkman, and H. Hilgenkamp, Nat Mater **5**, 556 (2006).
[3] N. Nakagawa, H.Y. Hwang, and D.A. Muller, Nat Mater **5**, 204 (2006).
[4] S. Thiel, G. Hammerl, A. Schmehl, C.W. Schneider, and J. Mannhart, Science **313**, 1942 (2006).
[5] N. Reyren, S. Thiel, A.D. Caviglia, L.F. Kourkoutis, G. Hammerl, C. Richter, C.W. Schneider, T. Kopp, A.-S. Rüetschi, D. Jaccard, M. Gabay, D.A. Muller, J.-M. Triscone, and J. Mannhart, Science **317**, 1196 (2007).
[6] A.D. Caviglia, S. Gariglio, N. Reyren, D. Jaccard, T. Schneider, M. Gabay, S. Thiel, G. Hammerl, J. Mannhart, and J.-M. Triscone, Nature **456**, 624 (2008).
[7] N. Reyren, S. Gariglio, A.D. Caviglia, D. Jaccard, T. Schneider, and J.-M. Triscone, Appl. Phys. Lett. **94**, 112506 (2009).
[8] A. Brinkman, M. Huijben, M. van Zalk, J. Huijben, U. Zeitler, J.C. Maan, W.G. van der Wiel, G. Rijnders, D.H.A. Blank, and H. Hilgenkamp, Nat. Mater. **6**, 493 (2007).
[9] A.J. Millis, Nat. Phys. **7**, 749 (2011).
[10] B. Kalisky, J.A. Bert, B.B. Klopfer, C. Bell, H.K. Sato, M. Hosoda, Y. Hikita, H.Y. Hwang, and K.A. Moler, Nat Commun **3**, 922 (2012).
[11] S. Banerjee, O. Erten, and M. Randeria, Nat. Phys. **9**, 626 (2013).
[12] M. Gabay and J.-M. Triscone, Nat. Phys. **9**, 610 (2013).
[13] J.-S. Lee, Y.W. Xie, H.K. Sato, C. Bell, Y. Hikita, H.Y. Hwang, and C.-C. Kao, Nat. Mater. **12**, 703 (2013).
[14] D.A. Dikin, M. Mehta, C.W. Bark, C.M. Folkman, C.B. Eom, and V. Chandrasekhar, Phys. Rev. Lett. **107**, 056802 (2011).
[15] L. Li, C. Richter, J. Mannhart, and R.C. Ashoori, Nat. Phys. **7**, 762 (2011).
[16] J.A. Bert, B. Kalisky, C. Bell, M. Kim, Y. Hikita, H.Y. Hwang, and K.A. Moler, Nat. Phys. **7**, 767 (2011).
[17] G. Herranz, F. Sánchez, N. Dix, M. Scigaj, and J. Fontcuberta, Sci Rep **2**, 758 (2012).
[18] A. Annadi, Q. Zhang, X. Renshaw Wang, N. Tuzla, K. Gopinadhan, W.M. Lü, A. Roy Barman, Z.Q. Liu, A. Srivastava, S. Saha, Y.L. Zhao, S.W. Zeng, S. Dhar, E. Olsson, B. Gu, S. Yunoki, S. Maekawa, H. Hilgenkamp, T. Venkatesan, and Ariando, Nat. Commun. **4**, 1838 (2013).
[19] Y.-L. Han, Y.-W. Fang, Z.-Z. Yang, C.-J. Li, L. He, S.-C. Shen, Z.-Z. Luo, G.-L. Qu, C.-M. Xiong, R.-F. Dou, L. Gu, C.-G. Duan, and J.-C. Nie. Submitted.
[20] A.F. Santander-Syro, O. Copie, T. Kondo, F. Fortuna, S. Pailhès, R. Weht, X.G. Qiu, F. Bertran, A. Nicolaou, A. Taleb-Ibrahimi, P. Le Fèvre, G. Herranz, M. Bibes, N. Reyren, Y. Apertet, P. Lecoeur, A. Barthélémy, and M.J. Rozenberg, Nature **469**, 189 (2011).
[21] Z. Wang, Z. Zhong, X. Hao, S. Gerhold, B. Stöger, M. Schmid, J. Sánchez-Barriga, A. Varykhalov, C. Franchini, K. Held, and U. Diebold, Proc. Natl. Acad. Sci. 201318304 (2014).
[22] T.C. Rödel, C. Bareille, F. Fortuna, C. Baumier, F. Bertran, P. Le Fèvre, M. Gabay, O. Hijano Cubelos, M.J. Rozenberg, T. Maroutian, P. Lecoeur, and A.F. Santander-Syro, Phys. Rev. Appl. **1**, 051002 (2014).
[23] V.L. Berezinskii, Sov. J. Exp. Theor. Phys. **34**, 610 (1972).
[24] J.M. Kosterlitz and D.J. Thouless, J. Phys. C Solid State Phys. **5**, L124 (1972).
[25] J.M. Kosterlitz and D.J. Thouless, J. Phys. C Solid State Phys. **6**, 1181 (1973).
[26] R. Bachelet, F. Valle, I.C. Infante, F. Sanchez, and J. Fontcuberta, Appl. Phys. Lett. **91**, 251904 (2007).
[27] K. Yoshimatsu, R. Yasuhara, H. Kumigashira, and M. Oshima, Phys. Rev. Lett. **101**, 026802 (2008).
[28] M. Sing, G. Berner, K. Goß, A. Müller, A. Ruff, A. Wetscherek, S. Thiel, J. Mannhart, S.A. Pauli, C.W. Schneider, P.R. Willmott, M. Gorgoi, F. Schäfers, and R. Claessen, Phys. Rev. Lett. **102**, 176805 (2009).
[29] J. Biscaras, N. Bergeal, A. Kushwaha, T. Wolf, A. Rastogi, R.C. Budhani, and J. Lesueur, Nat. Commun. **1**, 89 (2010).
[30] J. Biscaras, N. Bergeal, S. Hurand, C. Grossetête, A. Rastogi, R.C. Budhani, D. LeBoeuf, C. Proust, and J. Lesueur, Phys. Rev. Lett. **108**, 247004 (2012).
[31] M.R. Beasley, J.E. Mooij, and T.P. Orlando, Phys. Rev. Lett. **42**, 1165 (1979).
[32] B.I. Halperin and D.R. Nelson, J. Low Temp. Phys. **36**, 599 (1979).
[33] Z. Ovadyahu, Phys. Rev. Lett. **45**, 375 (1980).

[34] N.A.H.K. Rao, E.D. Dahlberg, A.M. Goldman, L.E. Toth, and C. Umbach, Phys. Rev. Lett. **44**, 98 (1980).

[35] K. Epstein, A.M. Goldman, and A.M. Kadin, Phys. Rev. Lett. **47**, 534 (1981).

[36] J.-H. She and A.V. Balatsky, Phys. Rev. Lett. **109**, 077002 (2012).

[37] D.J. Resnick, J.C. Garland, J.T. Boyd, S. Shoemaker, and R.S. Newrock, Phys. Rev. Lett. **47**, 1542 (1981).

[38] D.W. Abraham, C.J. Lobb, M. Tinkham, and T.M. Klapwijk, Phys. Rev. B **26**, 5268 (1982).

[39] A.M. Kadin, K. Epstein, and A.M. Goldman, Phys. Rev. B **27**, 6691 (1983).

[40] N. Cotón, M.V. Ramallo, and F. Vidal, Supercond. Sci. Technol. **24**, 085013 (2011).

[41] J.W. Ekin, Appl. Phys. Lett. **55**, 905 (1989).

[42] Y. Kozuka, M. Kim, C. Bell, B.G. Kim, Y. Hikita, and H.Y. Hwang, Nature **462**, 487 (2009).

[43] M. Ben Shalom, M. Sachs, D. Rakhmilevitch, A. Palevski, and Y. Dagan, Phys. Rev. Lett. **104**, 126802 (2010).

[44] M. Kim, Y. Kozuka, C. Bell, Y. Hikita, and H.Y. Hwang, Phys. Rev. B **86**, 085121 (2012).

[45] K. Gopinadhan, A. Annadi, Y. Kim, A. Srivastava, Ariando, and T. Venkatesan, ArXiv13117532 Cond-Mat (2013).

[46] Z.S. Popović, S. Satpathy, and R.M. Martin, Phys. Rev. Lett. **101**, 256801 (2008).

[47] P. Delugas, A. Filippetti, V. Fiorentini, D.I. Bilc, D. Fontaine, and P. Ghosez, Phys. Rev. Lett. **106**, 166807 (2011).

[48] W. Meevasana, P.D.C. King, R.H. He, S.-K. Mo, M. Hashimoto, A. Tamai, P. Songsiriritthigul, F. Baumberger, and Z.-X. Shen, Nat. Mater. **10**, 114 (2011).